# Intelligent Remote Control for TV Program based on Emotion in Arabic Speech

Mohamed MEDDEB, Hichem KARRAY, Adel.M.ALIMI
*Ecole Nationale des Ingénieurs de Sfax*
Research Group on Intelligent Machines REGIM
`Mohamed.meddeb@ieee.org`, `hichem.karray@ieee.org, Adel.alimi@ieee.org`

*Abstract*—Recommender systems for TV program have been studied for the realization of personalized TV Electronic Program Guides. In this paper, we propose automatic emotion Arabic speech recognition in order to achieve an intelligent remote control. In addition, the TV can estimate our interests and preferences by observing our behavior to watch and have a conversation on topics that might be interesting to us.

## I. INTRODUCTION

The speech signal conveys the message not only informative but also repository of information on personality and emotional state of the speaker.

The emotion is a motivational and adaptive response of an organism to the social environment. It is part of everyday life in.

Humans, Despite the long history of study of speech and that of emotion, relatively little work has been devoted to the analysis of emotional speech. This can be attributed to the formalism of modern science and the methodological problems in the study of vocal emotion. In formal linguistics, researchers have been concerned to describe the regularity of binary language, rather than the variability of the multipurpose floor, and they consider the change due to voice emotion as a random variable, non-routine, which does not deserve scientific analysis. In the psychology of emotion, researchers are primarily interested in demonstrating the nature of emotion (either pre-cognitive or cognitive-post) rather than to describe the vocal expression of emotion, facial or body. Moreover, most psychological studies of emotion are based on data of facial emotion rather than emotion voice data because the latter are more difficult to acquire than first. The formalism is also in the field of speech technology [1].

The recognition of emotions in speech has many useful applications. In the man-machine interfaces, robots can learn to interact with humans and to recognize human emotions. Robotic Pets, home lighting automation and interactive television (ITV), for example, should be able to understand not only the voice commands, but also other information, such as emotional health humans and modify their actions accordingly.

## II. PROBLEMATIC

, researchers increasingly realize the importance of knowledge on the emotional speech for the development of research in practice and in the theoretical . The overall mechanism of speech communication can be better understood by understanding the influence of emotion on the production and speech perception. The nature of emotion can be better understood by studying the relationship between the expressions voice, facial and bodily emotion. In speech technology, the addition of personal traits and emotional is essential to increase the naturalness of synthetic speech. The automatic speech recognition can greatly benefit from the development of the system that can recognize speech from different speakers in their various emotional states [2].

Our works is a study of emotional speech on the basis of data acquired from real situations. It aims to demonstrate how the emotion of the speaker is expressed in its natural speech in terms of acoustic cues and how the listener perceives it in different conditions of hearing. The analysis data consists of spontaneous speech excerpts, TV shows and reality, expressing the joy and sadness (whining voice) Tunisian speakers. Emotions predestined in this work are considered as the true emotions experienced by the speaker, as opposed to stylized emotions imitated by an actor. The choice of emotions actually experienced (not stylized) is based on the following consideration. Since the recording of vocal emotions from natural situations because of methodological problems and mental trauma, most studies are based on data from the emotions expressed by an actor following the experimenter's instructions about how voice. Although these theatrical emotions are supposed to be representative of those we experience in everyday life, both kinds of emotions are different yet at the emotional, motivational and expressive. The expression of emotion by the actor often takes an exaggerated form and involves a theatrical style, whereas the expression of 2 emotions by ordinary people in everyday social interaction takes rather a discrete and follows rules regulating exposure.

Considering the difference between emotion and the emotion experienced stylized and the scarcity of studies of emotion experienced, we decided to examine data emotional expressions natural, improvised from interviews. Although these talks have been broadcast through television, the emotions expressed in them are considered authentic compared to what the person felt.

This essay consists of a series of acoustic analysis on the emotional and neutral statements and a series of experiments on perceptual identification of emotion by listeners Tunisians

### III. RELATED WORK

#### A. *TV program recommendation*

Recommender systems for TV program have been studied for the realization of personalized TV Electronic Program Guides. To recommend appropriate TV programs, a user profile that reflects its preferences on the selection of TV programs should be estimated. The typical characteristic used to generate the user profile is the time to look, speech expressions, face and gesture detection and attributes of television programs. The characteristics of television programs, such as genres and performers (actors) are used to estimate the common characteristics of television programs, these attributes can be obtained from metadata, such as TV-Anytime, and IEPG (Internet Electronic Program Guide). The behavior of users to watch TV programs have also been studied.

#### B. *Emotions classification Algorithms*

As for the choice of emotions classifier there is no uniform a priori answer to the question of which classifier constitutes the best choice. The criterion for selecting an emotion classifier should be related to the task, in order to take into account the regularities of the problem, or of the geometry of the input feature space. Some classifiers are more efficient with certain type of class distributions, and some are better at dealing with many irrelevant features or with structured feature sets. One of the ways to compare the choice is to test the classifiers on the same large and representative database. Many classifiers have been tried for SER[1], and after Weka 1 has appeared it has become easy and straightforward. The most frequently used are Support Vector Machines SVM, Gaussian mixture models GMM, k nearest neighbor KNN, Mel-frequency Cepstral coefficients MFCC and Neural Networks NN [2].

A rapidly evolving area in pattern recognition research is the combination of classifiers to build the so-called classifier ensembles. For a number of reasons (ranging from statistical to computational and representational aspects) ensembles tend to outperform single classifiers. In the field of speech processing classifier ensembles have also proven to be adequate for the SER task [5].

### IV. PROPOSED RECOMMENDATION SYSTEM FOR TV PROGRAMS

The following diagram fig.1 depicts how we view the role of personality (user profile) and emotion as glue between perception, dialogue and expression. Perceptive data is interpreted on an emotional level by an appraisal model. This results in an emotion influence that determines, together with the personality what the new emotional state and mood will be. An intelligent agent uses the emotional state, mood and the personality to create behavior.

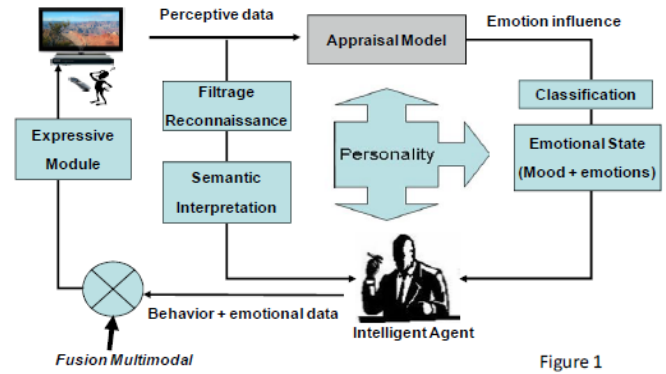

Fig 1. System proposed

To identify and decide on human behaviors convey in his speech, we have processes in stages. Processing sounds and audio segmentation sequences already stored in the speech database. Selecting and extracting relevant features. These latter parameters will forge the inputs classification system for emotions.

In this work we will treat the fundamentals phases of the steps of automatic recognition emotion.

### V. AUTOMATIC EMOTION RECOGNITION

Automatic emotion recognition of speech can be viewed as a pattern recognition problem [1]. The results of emotion recognition are characterized by: a) the features that are involved in the speaker's emotional state, b) the type of required emotions; c) the type of classifier used in the experiments and finally d) the database used for training and testing the classifier.

We develop in this section the key to a system of automatic emotion recognition from speech in real time elements. Support work is based on the basis of speech data, recorded to this effect in real and spontaneous conditions, and evaluated by a sample of the Tunisian population. In order to extract the features, we select energy; pitch, LPCC, MFCC that influences and better serve the emotional states. For classification, we adopted the Support Vector Machines (SVM) framework for Supervised Learning.

### A. Arabic Emotional Speech Database

Almost all the existing emotional speech databases have some limitations for assessing the performance of proposed emotion recognizers. Some of the limitations of emotional speech databases are briefly mentioned:

1. Most speech emotional databases do not well enough simulate emotions in a natural and clear way. This is evidenced by the relatively low recognition rates of human subjects. In some databases (see [94]), the human recognition performance is as low as about 65%.

2. In some databases such as KISMET, the quality of the recorded utterances is not so good. Moreover, the sampling frequency is some what low (8 kHz).

3. Phonetic transcriptions are not provided with some databases such as BabyEars [120]. Thus, it is difficult to extract linguistic content from the utterances of such databases.

A good database is as important as the desired results. There are different databases created by speech processing community with the help of professional actors which is widely used in research work. The results are uncompromising though the emotions are acted rather than spontaneous or natural. The famous databases are The Danish Emotional Speech Database (DES), and The Berlin Emotional Speech Database (BES), as well as The Speech under Simulated and Actual Stress (SUSAS) Database. DES and BES are representative for the early databases in the nineties but still serve as exemplars for acted emotional databases. For English, there is the 2002 Emotional Prosody Speech and Transcripts acted database available.

All of databases do not address the Arab emotional specificity and behaviour. Traditionally Arabic consonants are characterized by a rich and a poor vocalism. Arabic is a language which can be particular that differs from other language: the diversity of dialects in the Arab world, as well as the plurality of registers and varieties of language in different contexts of communication.

The emotion in speech depends on spoken language and especially the origin of the human being. in a state of emotion neutral descriptors varies from one person to another and from one social milieu to another. The reaction in a normal state of a citizen Mediterranean southern and the northern Mediterranean does not produce the meme expression and behaviour; say that in the case of other emotional state.

To explore this difference in behaviour, we propose to create a database Tunisian dialect sound. The database it's composed primarily of sound passages and isolated word (Table 1) recorded by actors where the ages, sex and region are different.

**Table 1:** Utterances to be recorded

| كلمة Words | نطق الكلمة spelling | جملة Phrases | نطق الجملة Spelling |
|---|---|---|---|
| تفضّل | tfadhal | اسمع نقلّك | Ism3 nkollik |
| عسلامة | asslama | شوف قدامك | Chouf nkollik |
| شدّ | chid | فك علي | Fok aliya |
| خوذ | khouth | قداش عمرك | Kaddach omrik |
| هات | hât | صلي على النبي | Salli ala ennabi |
| اعطيني | a3tini | بارك الله فيك | Baraka allah fik |
| شبيك | chbik | لا اله الا الله | La ilaha illa allah |
| اقعد | ok3ad | اخرج عليّ | Okhrij alia |
| قوم | koum | شنو إسمك | Chnoua ismik |
| إجري | ijri | سّيب من يدّك | Saib min iddik |
| نقّص | nakkas | أقسم بالله | Oksom billah |
| قريب | karib | ازرب روحك | Isrib rouhik |
| بعيد | ba3id | السلام عليكم | Assalamou alikom |
| لاباس | labes | قول الحق | Koul alhak |

- **Recording of the Database**

In order to evaluate emotion in speech, we decided to record a different form of utterances (tab.1). First choice is very short utterance like "شدّ (chid)", second is sentences like "فك علي (fok alia)", the third one is passages of fluent speech as described. For the TUES database the following is recorded:

- 10 single words

- 10 sentences and

- 2 passages of fluent speech.

The utterances above are spoken by actors once for each of the six emotions. Band of age are: 12-18, 18-25, 25-40, 40-60.

- **Recording Conditions**

Sound recording studios will be in professional audiovisual production. A recording room with good acoustics, a professional audio mixer, reliable microphones without background noise and anti pop will be used. Syntheses and word (Tab.1) will be written in phonetics, in order to translate themes in several languages and dialects.

The sentences were designed to use for recording the seven emotions (Neutral, Anger, Surprise, Disgust, Fear, Happiness, and Sadness). This database contains speech 720 samples. The length of speech simples is up to 5 Seconds.

The spectral representation of the speech signal "by the emotion neutral" and the speech signal "by the emotion anger", clearly shows the difference of behaviour in its different states individuate emotional and psychological.

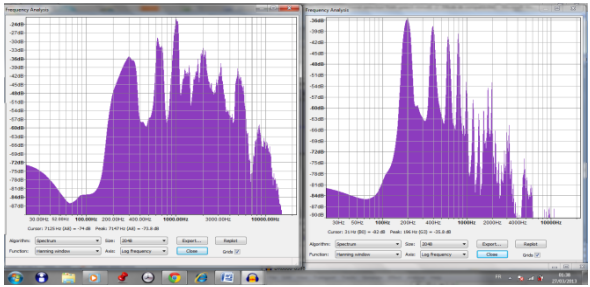

**Fig.2.a**  Neutral (لا)    **Fig.2.b**  Anger (لا)

The envelope of the signal is distinguished in both cases. The scale of frequency shows that the spectral in neutral state is richer in frequency than the angry one.

- **Performing the Listening Test**

A listening test was performed to test whether listeners could identify the emotional content of the recorded utterances.

The emotions was correct identified in 67% of the cases, ranging from 59% to 76%, see Table 2 .Surprise and happiness was often confused as well as neutral and sadness. The confusions between the emotions are shown in Table 2 where the emotions vertically are the emotions the actors tried to induce, and the emotions horizontally are the emotions interpreted by the listeners.

**Table 2:** *Confusions between the emotions for all listeners*

| Act / Listn | RESPONSE in % | | | | | |
|---|---|---|---|---|---|---|
| | Neu | Sup | Hap | Sad | Ang | Fea |
| Neu | 65.2 | 2.9 | 0.1 | 27.3 | 5.2 | 0.2 |
| Sup | 10.0 | 59.2 | 30.2 | 1.0 | 1.2 | 5.1 |
| Hap | 5.5 | 29.8 | 61.2 | 1.7 | 3.4 | 1.1 |
| Sad | 13.5 | 1.7 | 0.2 | 75.3 | 0.3 | 0.3 |
| Ang | 10.2 | 8.5 | 4.5 | 1.7 | 76.5 | 56.3 |
| Fea | 2.3 | 8.2 | 0.5 | 8.1 | 10.6 | 61.5 |

## I. Feature extraction

The speech signal contains a large number of information which reflects the emotional characteristics. Many researchers have proposed important speech features which contain emotion information, such as energy, pitch frequency [2], formant frequency [3], Linear Prediction Coefficients (LPC), Linear Prediction Cepstrum Coefficients (LPCC), Mel-Frequency Cepstrum Coefficients (MFCC) and its first derivative [4].

In our work, many features are used, such as energy, pitch, Linear Prediction Cepstrum Coefficients (LPCC) Mel-Frequency Cepstrum Coefficients (MFCC) [9].

### . Energy and related features

The Energy is the basic and most important feature in speech signal. We can obtain the statistics of energy in the whole speech sample by calculating the energy, such as mean value, max value, variance, variation range, contour of energy [7 [8]..

### . Pitch and related feature

We calculate the value of pitch frequency in each speech frame , and obtain the statistics of pitch in the whole speech sample. These statistical values reflect the global properties of characteristic parameters. Each Pitch feature vector has the same 19 dimensions as energy

### . Linear Prediction Cepstrum Coefficients (LPCC)

LPCC embodies the characteristics of particular channel of speech, and the same person with different emotional speech will have different channel characteristics, so we can extract these feature coefficients to identify the emotions contained in speech. The computational method of LPCC is usually a recurrence of computing the linear prediction coefficients (LPC), which is according to the all-pole model.

### . Mel-Frequency Cepstrum Coefficients (MFCC)

MFCC is based on the characteristics of the human ear's hearing, which uses a nonlinear frequency unit to simulate the human auditory system. Mel frequency scale is the most widely used feature of the speech, with a simple calculation, good ability of the distinction, anti-noise and other advantages]. In our research, we extract the first 12-order of the MFCC coefficients.

## II. Classification

Classification is another component of a speech emotion recognition system. In this research, we use the Support Vector Machines (SVM) framework for supervised learning [5] [8]. Here we use the LIBSVM implementation with the Radial Basis Function (RBF) kernel. Advantage of using RBF kernel is that it restricts training data to lie in specified boundaries. The RBF kernel nonlinearly maps samples into a higher dimensional space, so it, unlike the linear kernel, can handle the case when the relation between class labels and attributes is nonlinear. The RBF kernel has less numerical difficulties than polynomial kernel.

- **Training Models**

The Arabic Emotion database contains 720 speech files for five emotion classes. We choose three from it. Emotion classes sad, happy, neutral. We use both database, combine different features to build different training models, and analyse their recognition accuracy. Table 3 shows different combination of the features for the experiment.

**Table 3 :** Different combination of the features for the experiment.

| Training Model | Combination of Feature Parameters |
|---|---|
| Model 1 | MFCC+LPPC |
| Model 2 | MFCC+LPCC+Energy+Pitch |

## VI. EXPERIMENTS AND RESULTS

The performance of speech emotion recognition system is influenced by many factors, especially the quality of the speech samples, the features extracted and classification algorithm. As is shown at Table 4 the feature combination of MFCC+MEDC+Energy +Pitch shows a good performance.

**Table 4.** The Recognition Rate and Cross Validation

| Training Model | Features Combination | Cross Validation Rate | Recognition Rate |
|---|---|---|---|
| Model 1 | MFCC+LPCC | 88% | 80% |
| Model 2 | MFCC+LPCC+Energy+Pitch | 92% | 93 % |

The cross validation rate is as high as 92%, and the recognition accuracy rate is also around 93%. The feature of Energy and pitch plays an important role in Arabic speech emotional recognition.

## VII. CONCLUSIONS

Emotion recognition is used in more and more applications. In this study, the emotion recognition from continuous speech is realized. The application of ITV will makes it very easy the choice of television programs that meet user preferences.

The different combination of emotional characteristic features can obtain different emotion recognition rate, and the sensitivity of different emotional features.

As can be seen from the experiment, the emotion recognition rate of the system which only uses the spectrum features of speech is slightly higher than that only uses the prosodic features of speech. And the system that uses both spectral and prosodic features is better than that only uses spectrum or prosodic features. Meanwhile, the recognition rate of that use energy, pitch, LPCC MFCC and is slightly lower than that only use energy, pitch MFCC and LPCC features. This may be accused by feature redundancy.

To extract the more effective features of speech and enhance the emotion recognition accuracy is our future work. More work is needed to improve the system so that it can be better used in real-time speech emotion recognition.